\documentclass[aps,pre
,floatfix
,twocolumn
,groupedaddress
,amsmath
,showpacs
]{revtex4}
\usepackage{graphicx}
\usepackage{bm}
\usepackage{dcolumn}
\usepackage[geometry]{ifsym}
\begin{document}

%\preprint{}

\title{Studies of Mass and Size Effects in Three-Dimensional
  Vibrofluidized Granular Mixtures} 


\author{Peter E.~Krouskop}
%\email[]{krouskop@duq.edu}
%\thanks{}
\affiliation{Department of Chemistry and Biochemistry, Duquesne University,
Pittsburgh, PA 15282-1530} % \affiliation command applies to all authors since
%the last \affiliation command. The \affiliation command should follow the
\author{Julian Talbot}
\email[Author to whom correspondence should be addressed. ]{talbot@duq.edu}
%\thanks{}
\affiliation{Department of Chemistry and Biochemistry, Duquesne University,
Pittsburgh, PA 15282-1530}


\date{\today}

\begin{abstract}
We examine the steady state properties of binary systems of driven
inelastic hard spheres.  The spheres, which move under the influence
of gravity, are contained in a vertical cylinder with a vibrating
base. We computed the trajectories of the spheres using an
event-driven molecular dynamics algorithm.  In the first part of the
study, we chose simulation parameters that match those of
experiments performed by Wildman and Parker \cite{WP2002}. Various
properties computed from the simulation including the density profile,
granular temperature and circulation pattern are in good qualitative
agreement with the experiments.  We then studied the effect of varying
the mass ratio and the size ratio independently while holding the
other parameters constant. The mass and size ratio are shown to affect
the distribution of the energy.  The changes in the energy
distributions affect the packing fraction and temperature of each
component.  The temperature of the heavier component has
a non-linear dependence on the mass of the lighter component, while
the temperature of the lighter component is approximately
proportional to its mass. The temperature of both components is
inversely dependent on the size of the smaller component.
\end{abstract}

%\keywords{}
\maketitle


\section{Introduction \label{sec:intro}} 
 
Granular systems exhibit many properties that are different from
systems composed of elastic particles.  For example, driven granular
systems display standing and traveling waves
\cite{DFL1989,PB1993,MUS1995}, oscillons \cite{UMS1996}, heaping, and
convection \cite{ER1989, EJKKKN1995}.  In addition, granular mixtures
show size segregation \cite{RSPS1987} and steady-state kinetic
energies that are not equal for each component in the mixture
\cite{L1999}.  This departure from equipartition is not unexpected, but
it is one of the most striking differences between elastic and
inelastic systems. Understanding the properties of mixtures is
particularly important for granular systems since, unlike molecular
systems, they are never completely monodisperse.
 
Theoretical studies of granular systems have focused on two distinct
classes.  One consists of systems that are not driven or heated.  The
initial energy decays over time as a result of inelastic collisions.
During this ``cooling'' process there is a period during which the
density is homogeneous.  Several workers have presented
kinetic theories \cite{ML1998,GD1999,DHGD} and mean-field theories
based on Maxwell models \cite{MP2002} to describe the properties of
mixed granular systems in this homogenous cooling state.  In the other
class of systems, an energy source, such as a vibrating wall, is
present.  This leads to a non-equilibrium steady state that has been
studied by several researchers \cite{ML1998,MP2002a,BT2002,BTa}.  In
both classes the components are predicted to have different kinetic
energies, or granular temperatures, that depend on the mass, size,
and restitution coefficient of the grains \cite{DHGD,BT2002}.
 
Recently, two-dimensional \cite{FM} and three-dimensional systems
\cite{L1999,WP2002} of driven, granular mixtures have been studied
experimentally.  Losert \emph{et al.}, who first reported the
difference in granular temperature \cite{L1999}, observed that the
velocity distributions deviated from those observed in systems with
elastic collisions.  Feitosa and Menon studied density distributions
and granular temperature profiles in two-dimensional systems with and
without gravity \cite{FM}.  Wildman and Parker have studied the
convection patterns, density distributions, and temperature profiles
in three-dimensional systems \cite{WP2002}.  These studies determined
that the heavier particles are at a higher granular temperature than
the lighter particles.  In both two and three-dimensional
systems, the ratio of the temperatures varies as the relative
proportion of the heavy and light particles is changed.  The
temperature ratio, however, is independent of the inelasticity of the
particles \cite{FM,WP2002}.
 
While the experimental techniques employed in the studies cited above
have provided many useful insights into granular behavior, they cannot
easily isolate the effects of particle mass, size, and
inelasticity. Theoretical and computational methods are useful in this
respect.  Molecular dynamics simulations of granular mixtures can 
accurately reproduce the phenomena observed in
experiment \cite{TV2002,PCMP}, while providing information on the
effects of the individual properties mentioned above.  For example,
Paolotti \emph{et al.}  \cite{PCMP} and Barrat and Trizac \cite{BTb}
investigated the effects of rotation, mass ratio, and relative density
in two-dimensional vibrated systems.  Mixtures have also been studied
under two-dimensional shear flow conditions \cite{CH2002}.
Simulations of the homogeneous cooling state in two-dimensional
systems are consistent with experiment and theory
\cite{DFL1989,BTb,CH2002}.

It is important to stress that the conclusions drawn from simulations
of two-dimensional systems cannot necessarily be extended to
three-dimensions.  In particular, the system boundaries have a much
larger influence in three-dimensions, as recently demonstrated by
Talbot and Viot \cite{TV2002}. 

This paper presents a three-dimensional, event-driven, molecular
dynamics simulation of a mixture of inelastic hard-spheres.  The
simulation methodology is discussed in Section \ref{sec:modsim}
followed by a comparison to the available experimental results,
Section \ref{sec:wandp}.  Finally, the effects of isolated changes in
the mass ratio (Section \ref{sec:mr}) and size ratio (Section \ref{sec:sr})
on the energy distribution and component temperatures are examined.
 
\section{Model and Simulation \label{sec:modsim}} 
 
\begin{figure} 
\includegraphics[width=3.375in,keepaspectratio=true]{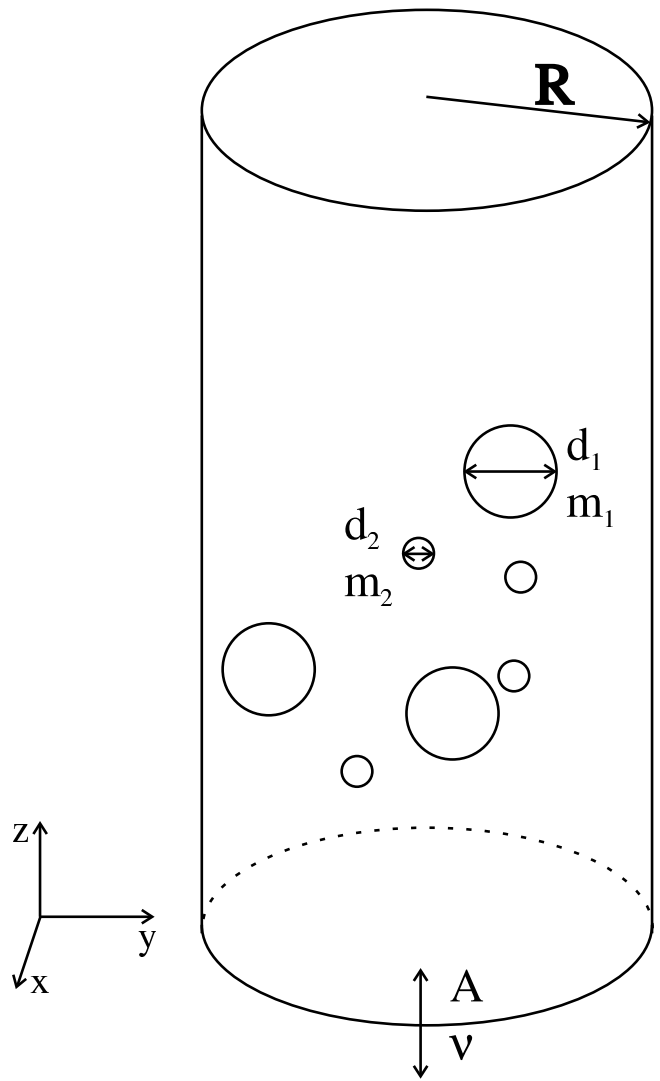} 
\caption{The three-dimensional system consisting of an infinitely tall 
  cylinder of radius $R$, and a mixture of hard spheres with sizes 
  $d_1$ and $d_2$ and masses $m_1$ and $m_2$.  The base of the 
  cylinder is shaken with a symmetric saw-tooth waveform with
  amplitude $A$ and frequency $\nu$. \label{fig:sys}}  
\end{figure}

The three-dimensional system \cite{TV2002} (Figure \ref{fig:sys})
contains a mixture of inelastic hard spheres in an infinitely tall
cylinder of radius $R$ under the influence of gravity.  The mixture is
composed of $n_1$ spheres of mass $m_1$ and diameter $d_1$ and $n_2$
spheres of mass $m_2$ and diameter $d_2$.  Energy is injected into the
system by means of the base of the cylinder, which vibrates in a
symmetric saw-tooth waveform of amplitude $A$ and frequency $\nu$.
 
Three kinds of collisions occur in the system: particle-particle,
particle-wall and particle-base. The post-collisional velocities
($\mathbf{v'_{\alpha,i}}$ and $\mathbf{v'_{\beta,j}}$) resulting from
a collision between a particle from component $\alpha$ and a particle
from component $\beta$ with masses $m_\alpha$ and $m_\beta$ and initial
velocities $\mathbf{v_{\alpha,i}}$ and $\mathbf{v_{\beta,j}}$,
respectively, are given by
\begin{subequations} 
\label{eqn:vel} 
\begin{equation} 
\label{eqn:vel1} 
\mathbf{v'_{\alpha,i}} = \mathbf{v_{\alpha,i}} -
\frac{m_\beta}{m_\alpha+m_\beta} \, (1+c) \lbrack
(\mathbf{v_{\alpha,i}} - \mathbf{v_{\beta,j}}) \cdot \mathbf{\hat{n}}
\rbrack \, \mathbf{\hat{n}}
\end{equation} 
\begin{equation} \label{eqn:vel2} 
\mathbf{v'_{\beta,j}} = \mathbf{v_{\beta,j}} +
\frac{m_\alpha}{m_\alpha+m_\beta} \, (1+c) \lbrack
(\mathbf{v_{\alpha,i}} - \mathbf{v_{\beta,j}}) \cdot \mathbf{\hat{n}}
\rbrack \, \mathbf{\hat{n}}
\end{equation} 
\end{subequations} 
The unit vector between the centers of the colliding particles is
$\mathbf{\hat{n}}$, and $c$ is the appropriate restitution
coefficient. While Equations \ref{eqn:vel1} and \ref{eqn:vel2}
conserve momentum, they imply an energy loss of
\begin{equation} 
\label{eqn:enloss} 
\Delta E = -\frac{1}{2}\mu (1-c^2) \lbrack (\mathbf{v_{\alpha,i}} - 
\mathbf{v_{\beta,j}}) \cdot \mathbf{\hat{n}} \rbrack ^2. 
\end{equation} 
where $\mu = m_\alpha m_\beta/\left(m_\alpha +m_\beta \right)$ is the
reduced mass of the particles involved in the collision.

In a binary mixture it is generally necessary to specify three
restitution coefficients for particle-particle collisions:\ $c_{11}$
and $c_{22}$ for intra-component collisions and $c_{12} (= c_{21})$
for inter-component collisions.  Several authors have reported on
random \cite{BTa, BTF2001} and velocity-dependent \cite{BSHP1996,
GSBMSS1998} restitution coefficients.  Luding and McNamara have
proposed a contact time model that also leads to a variable
restitution coefficient \cite{LM}.  For simplicity, we have chosen to
use a constant value for the restitution coefficients.  We further
assume that $c_{11} = c_{22} = c_{12} = c$.
 
Particle-wall collisions are governed by 
\begin{equation} 
\label{eqn:wall} 
\mathbf{v'_{\alpha,i}} = \mathbf{v_{\alpha,i}} - (1+c_{\alpha,w}) \,
(\mathbf{v_{\alpha,i}} \cdot \mathbf{\hat{r}}) \, \mathbf{\hat{r}} 
\end{equation} 
where $c_{\alpha,w}$ is the appropriate restitution coefficient for
component $\alpha$, and $\mathbf{\hat{r}}$ is the radial unit vector.
We assume that the restitution coefficient for collision with the wall
is constant for both species and $c_{1,w}=c_{2,w}=c_w$. Particle-base
collisions are governed by 
\begin{equation}  
\label{eqn:bottom} 
\mathbf{v'_{\alpha,i}} = \mathbf{v_{\alpha,i}} - (1+c_{\alpha,b}) \,
\lbrack (\mathbf{v_{\alpha,i}} - \mathbf{v_w}) \cdot \mathbf{\hat{k}} \rbrack
\, \mathbf{\hat{k}} 
\end{equation} 
where $c_{\alpha,b}$ is the appropriate restitution coefficient for
component $\alpha$, $\mathbf{v_w}$ is the velocity of the base at the
instant of collision, and $\mathbf{\hat{k}}$ is the unit vector in the
z-direction.  The restitution coefficients for collisions with the
base are also assumed to be constant and equal for both species
(i.e. $c_{1,b}=c_{2,b}=c_b$).
 
A phenomenon similar to inelastic collapse can be observed in these
simulations.  For certain ranges of the velocity a given particle will
collide repeatedly with the side wall.  As its energy is dissipated,
the particle approaches the wall ever more closely.  This is
accompanied by an increase in collision frequency that eventually
``freezes'' the simulation.  To prevent this phenomenon from
occurring, a small impulse is imparted to the particle toward the
center of the cylinder once its radial velocity falls below a certain
value.  This method has been used previously \cite{TV2002}, and the
threshold value was set such that the injected energy does not
discernibly influence the simulation output.  It is also possible for
a particle to come to rest on the base for a time corresponding to
$1/2$ a cycle of the vibration.  To avoid this possibility, the sign of the
z-component of the velocity for the particle is inverted when the
velocity of the colliding particle is found to match the velocity of
the base.  This condition was found to occur once every $12.5 \times
10^6$ collisions for the parameter values used in this paper.  Thus, this
method causes little perturbation in the simulation output.
 
We calculated a number of properties from the particle positions and
velocities generated by the simulation. The packing fraction
$\eta_{\alpha}$ for component $\alpha$ is defined as
\begin{equation} \label{eqn:pf} 
\eta_{\alpha} = \frac{n_{\alpha} \, v_{\alpha}}{V} 
\end{equation} 
where $n_{\alpha}$ is the number of particles of component $\alpha$ in
the volume element $V$, and $v_{\alpha}=\pi \, d^3_{\alpha}/6$ is the
volume of a particle of this component.  Another property that we
calculated is the kinetic energy or granular temperature,
$T_{\alpha}$, of each component using the following equations.
\begin{subequations} \label{eqn:temp} 
\begin{eqnarray} \label{eqn:temp2} 
T_{\alpha,x} = m_\alpha \, \langle v^2_{\alpha,x} \rangle \nonumber \\ 
T_{\alpha,y} = m_\alpha \, \langle v^2_{\alpha,y} \rangle \\ 
T_{\alpha,z} = m_\alpha \, \langle v^2_{\alpha,z} \rangle \nonumber 
\end{eqnarray} 
\begin{equation} 
\label{eqn:temp1} 
T_\alpha = \frac{(T_{\alpha,x} + T_{\alpha,y} + T_{\alpha,z})}{3} =
\frac{m_\alpha \, \langle \mspace{1mu} v_\alpha^2 \mspace{1mu} \rangle}{3} 
\end{equation} 
\end{subequations} 
The ``partial'' temperatures in the x-,y-, and z-directions are
$T_{\alpha,x}$, $T_{\alpha,y}$, $T_{\alpha,z}$, and the angular
brackets denote a time average over all particles of component
$\alpha$.
 
Our first objective was to model the experimental system studied 
by Wildman and Parker \cite{WP2002}.  We chose simulation 
parameters that correspond to those of the experiment, i.e., a cylinder 
of diameter 145mm that is shaken at 50 Hz with an amplitude of 1.74 
mm.  The acceleration due to gravity is taken as $g=9.81 m/s^2$.  The 
restitution coefficient for particle-particle collisions is $c = 
0.91$, for particle-wall collisions is $c_w = 0.68$, and for 
particle-base collisions is $c_b = 0.88$.   
 
We performed simulations in which we varied the relative proportions
of large and small particles while maintaining enough particles to
cover the base of the cylinder with a monolayer for comparison to
the experimental work of Wildman and Parker \cite{WP2002}.  We then
performed additional simulations to examine the effect of varying
the mass ratio $m_2/m_1$ and the size ratio $d_2/d_1$ independently.
In the discussion of these three studies, component 2 will always
refer to the smaller and/or lighter component.
  
The particles of each component were randomly placed in the cylinder
with random velocities.  We then equilibrated all systems for
approximately 5000 collisions per particle.  Data were collected over
$2.4 \times 10^4$ collisions per particle at intervals of
approximately 10 collisions per particle.  These data were then
averaged for each component to obtain a representation of the system
at steady state.
 
\section{Results and Discussion \label{sec:randd}} 
 
\subsection{Systems with varying composition \label{sec:wandp}} 
 
For the simulations discussed in this section, the size ratio of the
two components was set at $d_2/d_1 = 0.8$ and the mass ratio was set
at $m_2/m_1 = 0.512$.  Systems with the following compositions were
then studied: $n_1 = 525$, $n_2 = 270$; $n_1 = 350$, $n_2 = 540$; and
$n_1 = 175$, $n_2 = 810$.
 
\subsubsection{Velocity Field \label{sec:wandp:vf}} 
 
\begin{figure*} 
\includegraphics[height=4in,keepaspectratio=true]{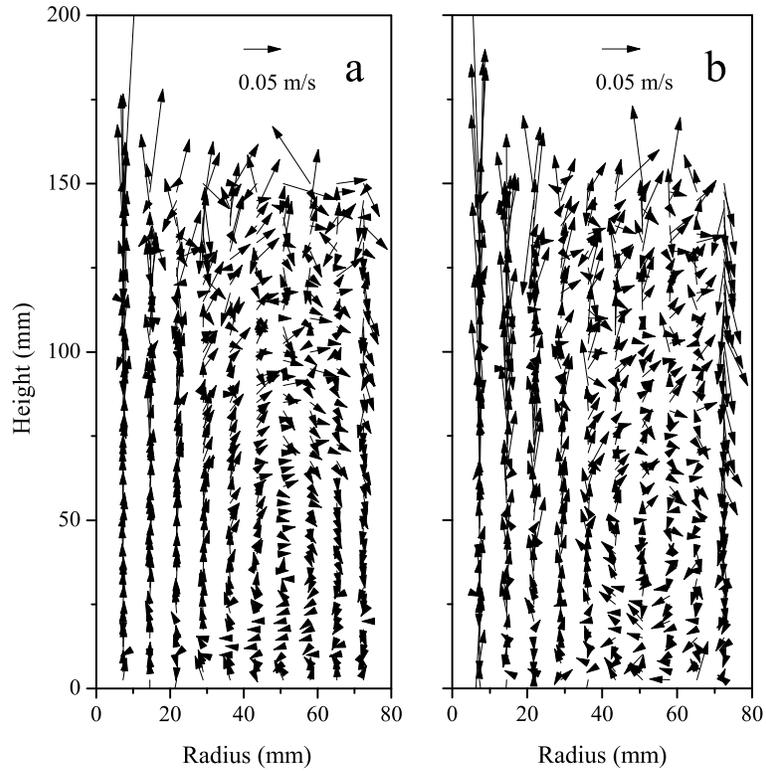}
\caption{Velocity field of (a) component 1 and (b) component 2.  Simulation 
conditions as follow: $n_1 = 525$, $n_2 = 270$, $d_2/d_1 = 0.8$, $m_2/m_1 = 
0.512$, all other conditions as given in the text. \label{fig:velfield}} 
\end{figure*}

The velocity fields of components 1 and 2 for a system with $n_1 =
525$ and $n_2 = 270$ particles are shown in Figures
\ref{fig:velfield}a and \ref{fig:velfield}b, respectively. It can be
seen that both components circulate in a pattern that rises in the
center of the cylinder and falls at the walls.  This convection
pattern has been observed previously in both experiment and simulation
\cite{WP2002,TV2002,WHP2001}.  The patterns shown here are very
similar to those reported by Wildman and Parker \cite{WP2002} with the
center of the convection for both species present at a radius of
approximately 50 mm and a height of approximately 40 mm.
 
\subsubsection{Packing Fraction \label{sec:wandp:pf}} 
 
\begin{figure*} 
\includegraphics[ height=4in, keepaspectratio=true]{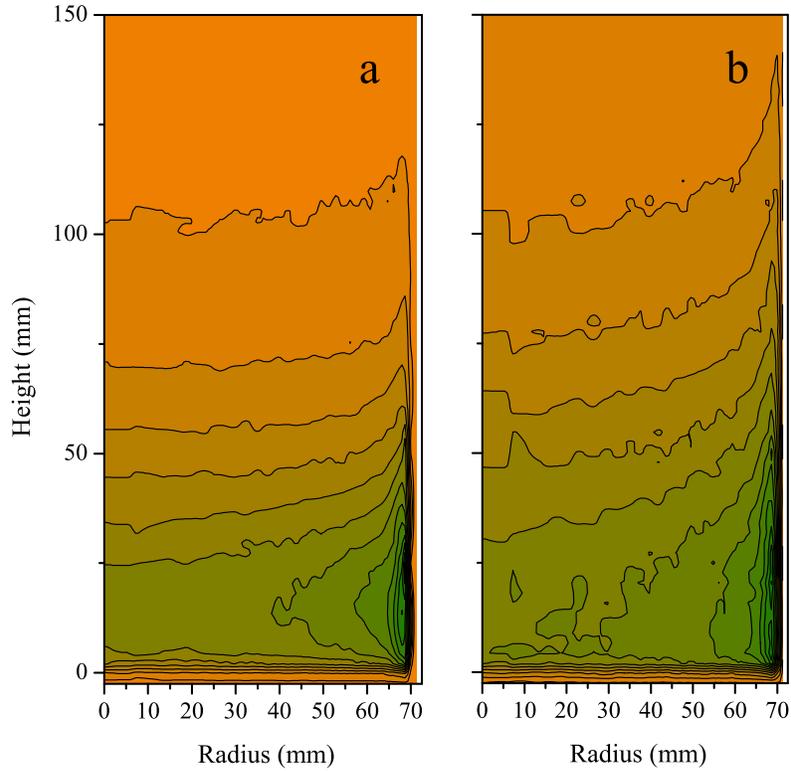}
\caption{Contour plots of the  packing fraction of (a) component 1 and
  (b) component 2.  Contours correspond to a packing fraction change
  of 0.002 in (a) and 0.0011 in (b).  Simulation conditions as given
  in Figure \ref{fig:velfield}.
\label{fig:wp:pf1}} \end{figure*}

The packing fraction, as a function of radius and height for each
component, is shown in Figure \ref{fig:wp:pf1}.  It can be seen that
there is a density gradient in both the radial and vertical directions
for both components, just as is observed in monodisperse systems
\cite{TV2002,WHP2001}.  The data from the simulation are qualitatively
similar to those observed in experiment except that the simulation
shows a higher concentration at the bottom of the cylinder near the
wall for both components.  It should be noted that while the maximum
density occurs at the same point for both components, component 2
obtains a greater height than component 1.  This trend is opposite
that observed for size segregation in weakly tapped systems where the
larger particles rise above the smaller ones \cite{RSPS1987}.
 
\begin{figure} 
\includegraphics[width=3.375in,
keepaspectratio=true]{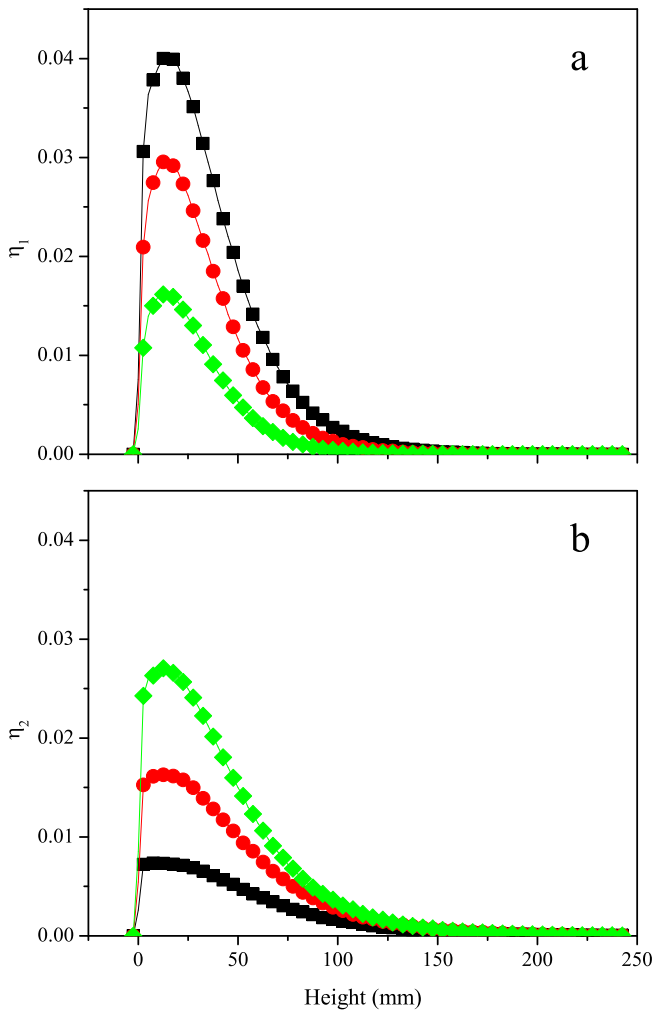}
\caption{Packing fractions of (a) component 1 and (b) component 2 for
  the compositions: \FilledSquare\ $n_1 = 525$, $n_2 = 270$;
  \FilledCircle\ $n_1 = 350$, $n_2 = 540$; \FilledDiamondshape\ $n_1 =
  175$, $n_2 = 810$.  Other simulation conditions as given in Figure
  \ref{fig:velfield}. \label{fig:wp:pf2}}
 \end{figure}

The radially averaged packing fractions as a function of height are
shown in Figure \ref{fig:wp:pf2}.  The three curves correspond to the
three relative fractions of component 1 and component 2.  In all
cases, the packing fraction of both components increases steeply at
small heights, reaches a maximum and decays at large heights.  The
packing fraction of component 1 decreases and it increases for
component 2 as the relative amount of each component is
changed.  The changes in the relative fractions only affects the
magnitudes of the packing fractions.  There is no noticeable variation
in the details of the packing fraction profiles (i.e. the position of
the maximum, the rate of decay, etc.) as the relative amounts of the
components are changed.

%reduced (Figure
%\ref{fig:wp:pf2}a).  The packing fraction of component 2 increases as
%its relative fraction is increased (Figure \ref{fig:wp:pf2}b). also
%shows that the packing fractions of component 1 
%decrease in area as the fraction of large particles decreases.  The
%decrease in the area is due to the reduced number of particles.
%Figure \ref{fig:wp:pf2}b shows that the packing fractions of component
%2 increase in area as the fraction of large particles decreases.  The
%increase in area is caused by the increase in the number of small
%particles within the system.
 
\subsubsection{Temperature \label{sec:wandp:t}} 
 
\begin{figure*} 
\includegraphics[height=4.0in, keepaspectratio=true]{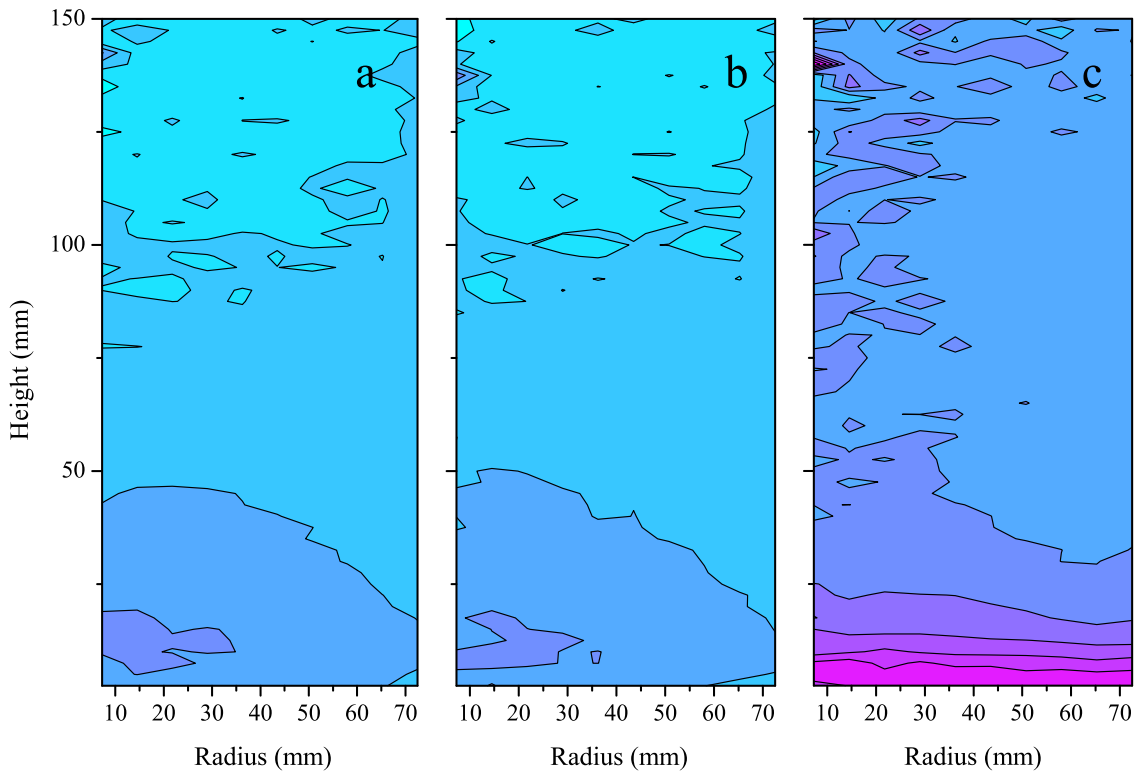}
\caption{Contour plots of the temperature in the (a) x-direction, (b)
  y-direction, and (c) z-direction for component 1.  Contours
correspond to a change of $11.5 \; \mu$J.  Simulation conditions as
given in Figure \ref{fig:velfield}. \label{fig:wp:t1}} \end{figure*}

The granular temperature of each component is also studied as a
function of the relative proportions.  Figure \ref{fig:wp:t1} shows
contour plots of the granular temperature of component 1 in the x-
(Figure \ref{fig:wp:t1}a), y- (Figure \ref{fig:wp:t1}b), and z-
(Figure \ref{fig:wp:t1}c) directions for the system presented in
Figure \ref{fig:velfield}.  The symmetry evident in the x- and
y-directions is produced by the unbiased introduction of energy into
these directions by particle-particle collisions.  These two partial
temperatures decay rapidly in both the radial and vertical dimensions
from a maximum near the center of the cylinder, close to the vibrating
base. The z-direction, however, is different because of the bias
introduced by the vibrating base.  This partial temperature decays in
the vertical dimension with very little radial dependence at small
heights.  At larger heights, the center of the cylinder is slightly
warmer than the surrounding area, as would be expected given the
toroidal flow profile presented in Figure \ref{fig:velfield}.

%the result of particle-particle collisions introducing energy in these
%directions.  The collisions do not show a preference for injecting
%energy in either the x- or y-direction within the system.  The
%z-direction, however, is different because of the vibrating base.  It
%can be seen that the temperature in the x- and y-directions decay as
%both radius and height increase.  However, the temperature in the z
%direction has only a slight dependence on the radius, while decaying
%as height increases.  It can also be seen that the radial decay
%pattern for the three directions is different.  The energy in the x-
%and y-directions decays quickly at all radii.  The z-direction,
%however, shows a slower decay rate for the small radii than for the
%large radii.  It is also possible to observe a small minimum in the
%temperature for the small radii as the height increases.  This
%phenomenon can be better seen in Figures \ref{fig:wp:t2}a and
%\ref{fig:wp:t2}b.
 
\begin{figure} 
\includegraphics[width=3.375in, keepaspectratio=true]{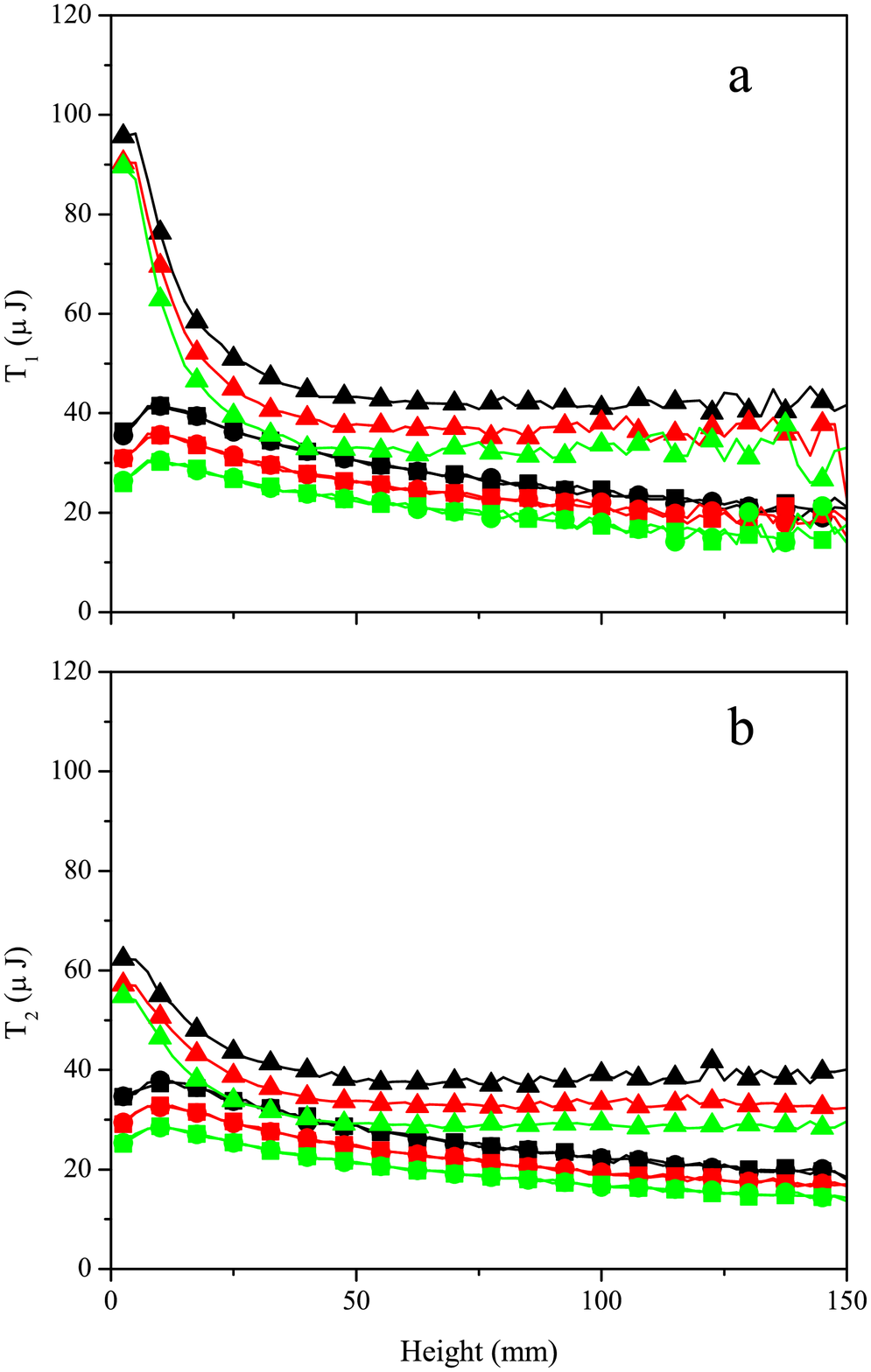}
\caption{Temperatures in the \FilledSquare\ x-, \FilledCircle\ y-, and 
\FilledTriangleUp\ z-directions of (a) component 1 and (b) component
2.  The three spacial directions are shaded the same for each system.
Simulation conditions as follows: $n_1=525$, $n_2=270$ (top curves);
$n_1=350$, $n_2=540$ (middle curves); $n_1=175$, $n_2=810$ (bottom
curves).  \label{fig:wp:t2}} 
\end{figure}

Figure \ref{fig:wp:t2} shows the height dependence of the radially
averaged, partial granular temperatures of each component.  It can be
seen that the height profiles of the temperature are similar for both
components at all relative proportions.  The temperatures in the x-
and y-directions show the formation of a maximum for both components
as height increases.  These maxima occur very close to the height that
corresponds to the maximum in packing fraction.  The increase in the
partial temperatures can then be attributed to an increase of
particle-particle collisions that inject energy into the x- and
y-directions, increasing the corresponding temperatures.  Figure
\ref{fig:wp:t2} also shows that the temperature in the z-direction is
larger than that in the other two directions.  The minimum observed as
height increases has been predicted by Brey \emph{et al.} for systems
in which the particles do not interact with a top barrier, but are
under the influence of gravity \cite{BR-MM2001}.  Wildman \emph{et
al.} observed the minimum the experimental systems
\cite{WHP2001,WHP2001a}, and Ramirez and Soto presented a
hydrodynamic theory that addresses this phenomenon \cite{RS}.
 
Figures \ref{fig:wp:t2}a and \ref{fig:wp:t2}b also show that the
temperatures of the two components change as the relative proportion
of the two components changes.  The temperature in the three spatial
directions decreases as the relative proportion of the larger
particles is decreased.  As the fraction of component 1 decreases, the
number of collisions with the smaller particles increases, causing the
temperature of the larger particles to decrease.  Also, the
temperature of the entire system decreases because component 2 does
not gain as much kinetic energy from the base since it is lighter than
component 1.  The decrease in the temperature, however, does not
change the height at which the extrema in $T_x$, $T_y$ and $T_z$ are
observed.  Only the magnitude of the measured temperature appears to
change in these systems (Figure \ref{fig:wp:t2}a and
\ref{fig:wp:t2}b).
 
While the trends observed in the temperature of the two components are
similar, there are differences between the two components.  The
temperatures of component 1 (Figure \ref{fig:wp:t2}a) are greater than
those of component 2 (Figure \ref{fig:wp:t2}b) in all the systems, an
effect particularly pronounced in the z-direction.  The temperatures
in the x- and y-directions differ only slightly.  It is difficult to
determine whether the differences in temperature between the two
components are dominated by the differences in size or mass from these
data.  Thus, we conducted further simulations to determine the
individual effects of mass and size.
 
\subsection{Systems with varying mass ratio \label{sec:mr}} 
 
To determine the effects of particle mass, we simulated systems with
mass ratios $m_2/m_1 =$~0.01, 0.125, 0.25, 0.5, and 1.0 at constant
size, $d_2/d_1 = 1.0$, and relative fraction $n_2/n_1 = 1.0$ and with
a total $n_1+n_2=1050$ particles in the system.
 
\subsubsection{Energy Distribution \label{sec:mr:energy}} 

\begin{figure} 
\includegraphics[width=3.375in, keepaspectratio=true]{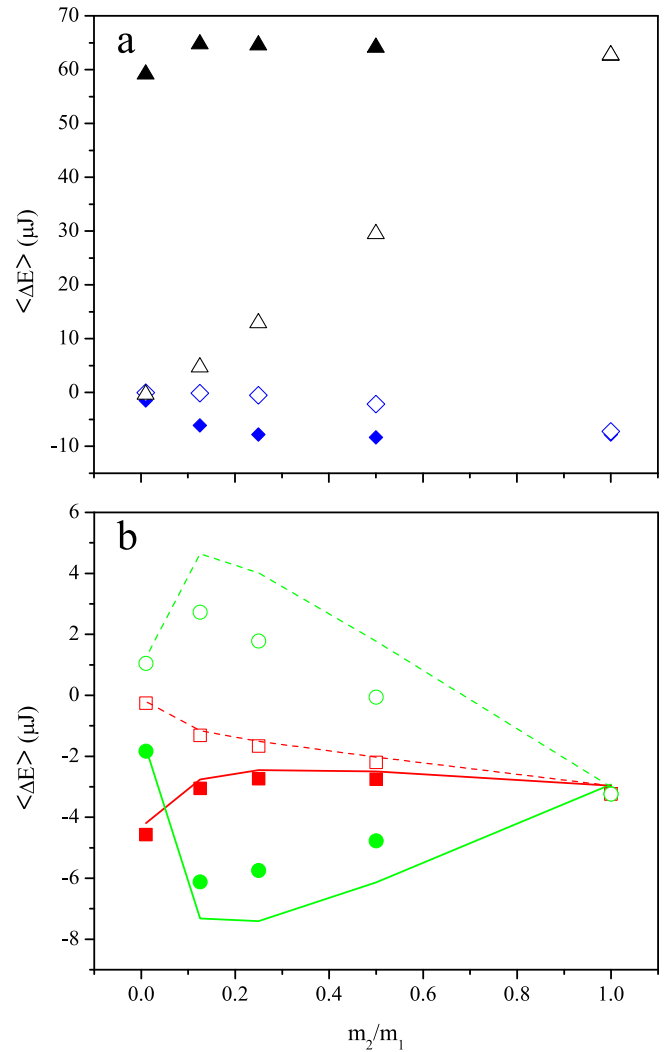}
\caption{Average energy change per particle per collision for
  (a) particle-boundary collisions and (b) particle-particle
  collisions as the mass ratio is changed.  Component 1 is represented
  by the closed symbols and  component 2 is represented by the open
  symbols.  The different kinds of collisions are: 
  \FilledSquare \ intra-component, particle-particle ($\langle \Delta
  E_1 \rangle_1$ and $\langle \Delta E_2 \rangle_2$), \FilledCircle \
  inter-component, particle-particle ($\langle \Delta E_1 \rangle_2$
  and $\langle \Delta E_2 \rangle_1$), \FilledDiamondshape \
  particle-wall, and \FilledTriangleUp \ particle-base. The solid
  (dashed) lines are the energy loss for component 1 (component 2) as
  calculated by Equation \ref{eqn:enloss1} for intra-component (same
  shade as the squares) and inter-component (same shade as the
  circles), particle-particle collisions. \label{fig:mr:energy}}
\end{figure}

We expect changes in the mass ratio to affect the exchange
of energy, and hence the temperature, of each component. 
We therefore computed the average change in energy, $\langle \Delta
E_\alpha \rangle_\beta$, of a particle of component $\alpha$ 
resulting from collisions with particles of component $\beta$.  We
examined the energy exchanges resulting from particle-particle,
particle-wall and particle-base collisions. The data we collected
from the simulations for each kind of collision are presented in
Figure \ref{fig:mr:energy}.  

As expected, both components lose energy on collision with the wall,
while they experience a net energy gain on collision with the base
(\ref{fig:mr:energy}a). Both components also lose energy from
intra-component, particle-particle collisions (squares in
\ref{fig:mr:energy}b).  The inter-component, particle-particle
collisions (circles in \ref{fig:mr:energy}b), however, show trends
that are not intuitively obvious. Component 1 shows a loss of energy
for all mass ratios, while component 2 shows that there may be a loss
or gain of energy depending on the mass ratio. We obtained a
theoretical estimate for this quantity by assuming each component has
a Maxwell-Boltzmann velocity distribution but with a temperature
specific to the component. The details of the derivation are presented
in the Appendix. The average energy loss for a particle of component
$\alpha$ resulting from collisions with particles of component $\beta$
and average component temperatures (kinetic energies) of $T_\alpha$
and $T_\beta$ is
%\begin{widetext} 
\begin{equation} 
\label{eqn:enloss1} 
\left< \Delta E_{\alpha} \right>_\beta = k_B \mu \left( 1+c \right)  
\left( \left(1+c \right) \frac{T_\beta m_\alpha + 
  T_\alpha m_\beta }{m_\alpha \left(m_\alpha+m_\beta 
  \right)}-2\frac{T_\alpha}{m_\alpha} \right)  
\end{equation}
where $k_B$ is the Boltzmann constant.  The total average energy loss
per collision between particles of components $\alpha$ and $\beta$
with temperatures of $T_\alpha$ and $T_\beta$ is  
\begin{equation} 
\label{eqn:enloss2} 
\left< \Delta E_{\alpha} \right>_\beta+\left< \Delta E_{\beta} \right>_\alpha =
-k_B \left( 1-c^2 \right)
\frac{T_\beta m_\alpha + T_\alpha m_\beta}{m_\alpha+m_\beta}
\end{equation} 
%\end{widetext} 
Equation \ref{eqn:enloss2} indicates that for inelastic collisions,
$c<1$, the total energy of the colliding pair always decreases, even
if the temperatures of the two components are different. For equal
temperatures, Equation \ref{eqn:enloss1} shows that this is also true
for the individual energies of the components. If, however, the
temperatures are different, it is possible for the energy of the
lighter component to \emph{increase} on average due to collisions with
the heavier component. 

We calculated values of $\langle \Delta E_{\alpha} \rangle_\beta$ for
inter-component and intra-component, particle-particle collisions for
each component using Equation \ref{eqn:enloss1}.  The masses of the
particles and the restitution coefficient are set by the input
parameters, but the temperatures of each component are not known
\emph{a priori}.  Therefore, the temperature was obtained from the
simulation output for each mass ratio.  The results from Equation
\ref{eqn:enloss1} are presented in Figure \ref{fig:mr:energy}b along
with the results obtained from the simulation.  It can be seen that
the average energy changes calculated for the two kinds of
particle-particle collisions compare favorably with those obtained
from the simulation.  There is better agreement for intra-component
collisions because there is no temperature or mass difference between
the colliding particles.  The predicted energy changes for
inter-component collisions show the same trend that is observed in the
simulation results, but the magnitude of the change is
incorrect. Specifically, the equation overestimates the change in
energy resulting from inter-component, particle-particle
collisions. This error probably arises from the inhomogeneities in the
particle density and temperature (see Figures \ref{fig:wp:pf1} and
\ref{fig:wp:t1}), which are not accounted for in the Maxwell-Boltzmann
distribution.

More generally, we note that as the mass ratio decreases, the energy
change associated with any collision also decreases.  This variation
in the energy change will affect the bulk properties observed for
these systems, such as the packing fraction and the temperature
discussed below. 
 
\subsubsection{Packing Fraction \label{sec:mr:pf}} 
 
\begin{figure} 
\includegraphics[width=3.375in, keepaspectratio=true]{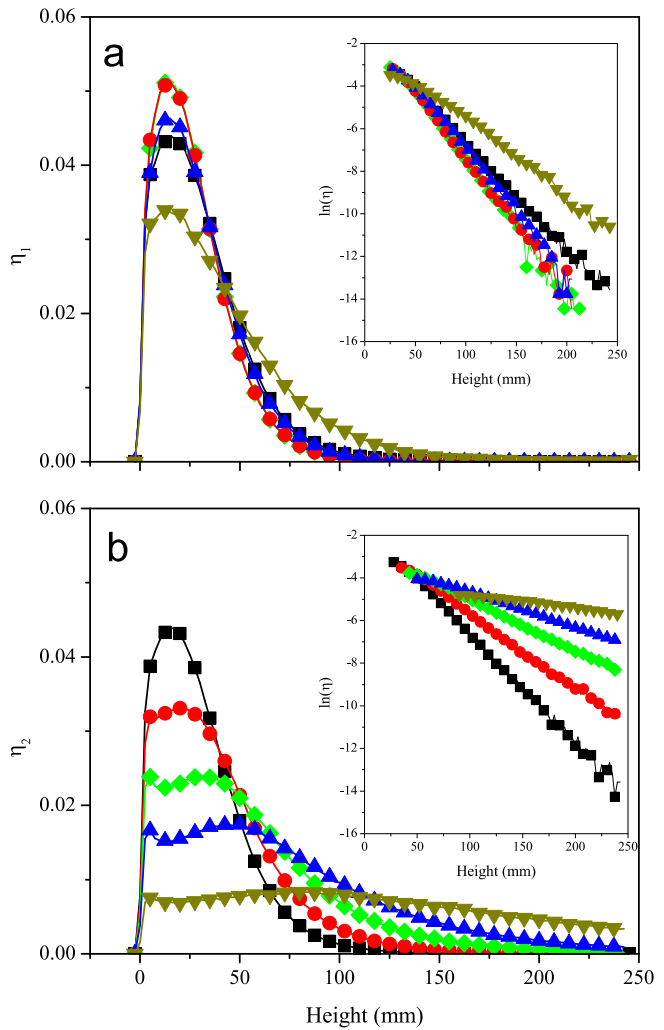}
\caption{Packing fraction of (a) component 1 and (b) component 2 for
  mass ratios of \FilledSquare \   $m_2/m_1 =1.0$,
  \FilledCircle \ $m_2/m_1 = 0.5$, \FilledDiamondshape   \ $m_2/m_1 =
  0.25$, \FilledTriangleUp \ $m_2/m_1 = 0.125$, and
  \FilledTriangleDown \ $m_2/m_1 = 0.01$.  All other simulation
  parameters are as given in the text.  The inset shows the natural
  log of the trailing edge of the packing fraction for each system. 
  \label{fig:mr:pf}}  
\end{figure}

The packing fraction of each component as a function of height is
presented in Figure \ref{fig:mr:pf}.  The general behavior is the same
for both components, and is similar to that already discussed in
Section \ref{sec:wandp:pf}.  The packing fraction of component 1
varies little with mass ratio changes. In particular, the maximum
density is at the same height for all the systems examined. At greater
heights, however, component 1 condenses and then expands as the mass
of component 2 decreases.  This phenomenon is more clearly visible in
the insert of Figure \ref{fig:mr:pf}a, which shows a linear
relationship between $\ln(\eta)$ and the height.  At large altitudes,
the slope of the lines in the insert increase as component 1 condenses
and decrease as component 1 expands in the system.  This behavior
corresponds to the increased energy loss that is observed for
inter-component particle-particle collisions (see Figure
\ref{fig:mr:energy}).
 
The packing fraction profiles of component 2 undergo a much more
significant change as the mass ratio is decreased. Specifically,
Figure \ref{fig:mr:pf}b indicates a steady depletion of this component
from around the maximum, with a compensating increase at large
altitudes, as the particles expand into the upper reaches of the
cylinder. For mass ratios of 0.25 and below, two local maxima are
present.  These are most distinct for the systems in which component 2
has a net gain in energy due to collisions with particles of component
1 (mass ratios of 0.125 and 0.01).  Thus, the two maxima are formed as
the particles of component 2 try to separate from component 1.  The
increase in the energy forces the particles of component 2 toward the
base and toward higher altitudes.  This is what is observed in the
packing fractions shown in Figure \ref{fig:mr:pf}b with one maximum
very close to the base, and one maximum that increases in altitude
as the mass ratio decreases.  The plots of $\ln(\eta)$, shown in the
insert, display a steady decrease in the slope as the mass ratio
decreases.
 
\subsubsection{Temperature \label{sec:mr:t}} 
 
\begin{figure} 
\includegraphics[width=3.375in, keepaspectratio=true]{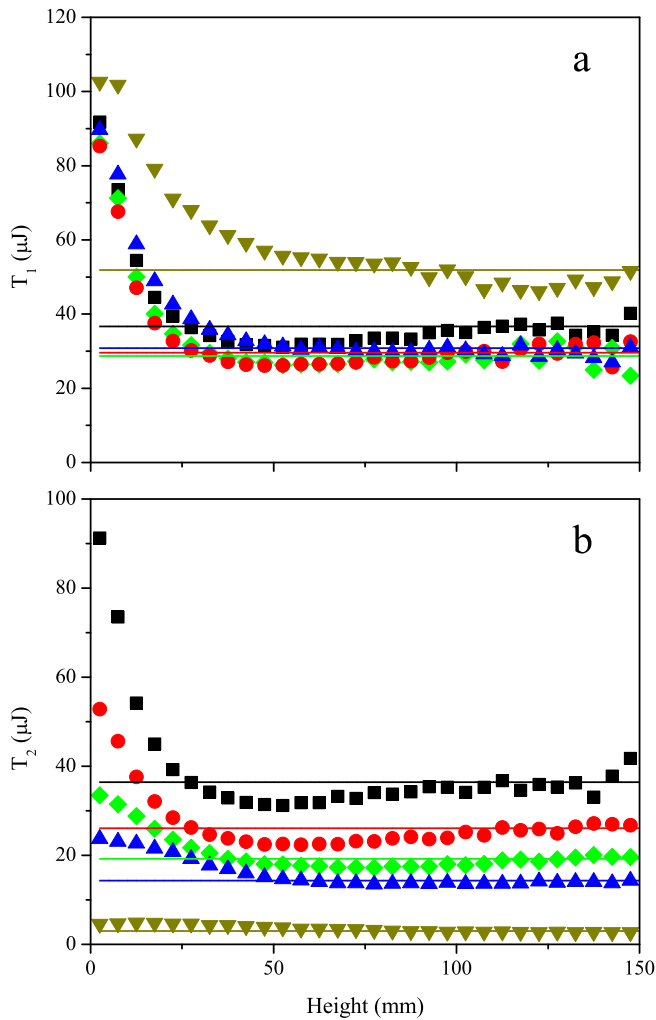}  
\caption{The temperature in the z-direction of (a) component 1 and
(b) component 2 for mass ratios of \FilledSquare~$m_2/m_1 = 1.0$,
  \FilledCircle~$m_2/m_1 = 0.5$, \FilledDiamondshape~$m_2/m_1 = 0.25$,
  \FilledTriangleUp~$m_2/m_1 = 0.125$, and
  \FilledTriangleDown~$m_2/m_1 = 0.01$.  The temperature calculated
  from Equation \ref{eqn:slope} is shown as a line corresponding to
  the data points of the same shade.  \label{fig:mr:t}}
\end{figure}

The effect of mass ratio on the temperature in the z-direction for
components 1 and 2 is shown in Figures \ref{fig:mr:t}a and
\ref{fig:mr:t}b, respectively.  The minimum in the temperature is
obvious for both particles.  It is also easily seen that the
temperature in the z direction of component 1 goes through a minimum
as the mass ratio of the two components decreases.  The temperatures
in the x- and y-directions (not shown) also follow the same trend.
This indicates that total temperature for component 1 goes through a
minimum as the mass ratio is decreased.  The changes in the
temperature coincide with the changes observed in the packing fraction
(Figure \ref{fig:mr:pf}a) and the energy changes for the different
kinds of collisions (Figure \ref{fig:mr:energy} ).  This implies that
the changes in the velocities of the particles of component 1 affect
the temperature, just as expected from Equations \ref{eqn:temp}.  As
shown by Brey \emph{et al.} \cite{BR-MM2001} and Warr \emph{et al.}
\cite{WHJ1995}, a relationship exists between the temperature and the
packing fraction in a single component system. Extending their
results to a multi-component system, we obtain
\begin{equation} 
\frac{d \ln (\eta_\alpha)}{d z}\sim -\frac{m_\alpha g}{k_B T_\alpha}
\label{eqn:slope} 
\end{equation} 
This equation holds for high altitudes and restitution coefficients
close to 1.  Thus, a limiting temperature can be calculated for each
system using the data presented in the insert of Figure
\ref{fig:mr:pf}a.  The results, plotted as horizontal lines in Figure
\ref{fig:mr:t}b, correspond well with the asymptotic temperatures.
This indicates that the decaying edge of the packing fraction is a
good indicator of the temperature at those altitudes.

We observed trends for component 2 that are very different from those 
just discussed for component 1.  Figure \ref{fig:mr:t}b shows that the 
temperature decreases as the mass ratio decreases.  The decrease is 
expected since temperature is directly related to the mass of the 
particle (equations (\ref{eqn:temp})).  Thus, the lighter particles 
will have a lower temperature than the heavier particles.  Component 2 
exhibits a minimum as the height increases, just as in the case of 
component 1.  However, the minimum becomes shallower as the 
temperature decreases.  Figure \ref{fig:mr:t}b also shows the 
temperature obtained from the packing fraction using Equation 
\ref{eqn:slope}.  Again, we find that there is good agreement 
between the temperature calculated by Equation \ref{eqn:slope} and 
that calculated from Equations \ref{eqn:temp} for large heights.  
 
\subsection{Effect of varying size ratio \label{sec:sr}} 
 
Finally, we studied the effects of particle size by simulating
systems with size ratios of $d_2/d_1 =$~1.0, 0.8, 0.5, and 0.1.  The
mass ratio was held constant at $m_2/m_1 = 1.0$,
and the relative fraction was held constant at 
$n_2/n_1 = 1.0$, with $n_1+n_2=1050$.  The changes observed
in the energy distributions, the packing fraction and the partial
temperatures are presented and discussed below.
 
\subsubsection{Energy Distribution \label{sec:sr:energy}} 

Changes in particle size results in changes in the mean free path
\cite{BR-MM2001} and the pair correlation function at contact
\cite{GD1999,BT2002,BTa}.  These changes affect the particle
velocities by changing the number of collisions that a particle
experiences in a given amount of time.  Thus, changes in the size
ratio are expected to result in changes in the distribution of the
energy, just as observed for the mass ratio.

\begin{figure} 
\includegraphics[width=3.375in, keepaspectratio=true]{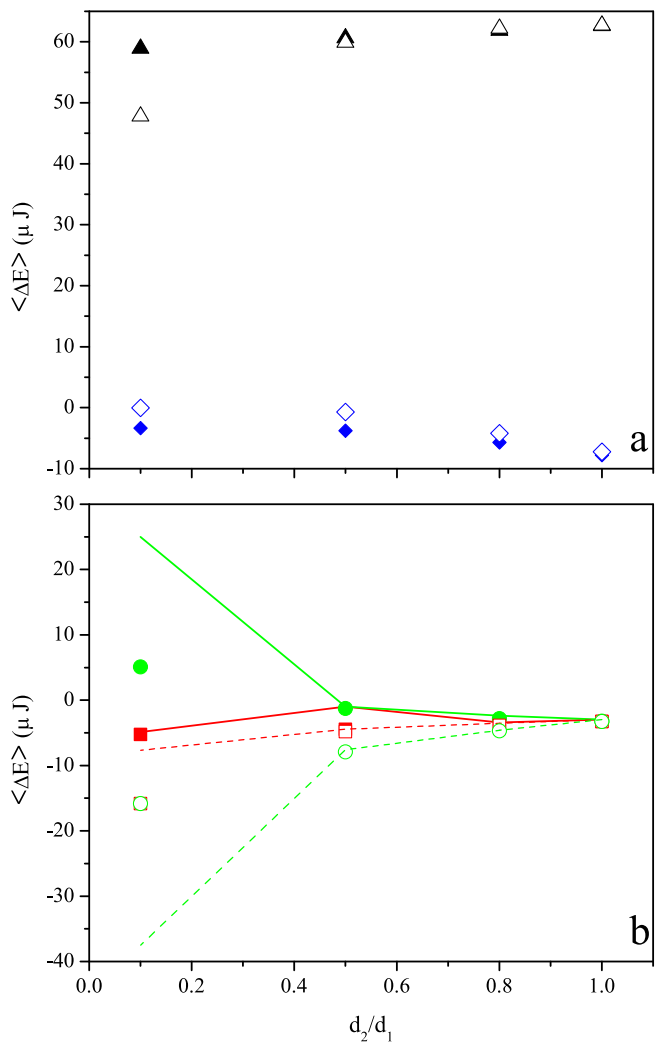}
\caption{Average energy change per particle per collision for (a)
  particle-boundary collisions and (b) particle-particle collisions as
  the size ratio is changed.  Component 1 is represented by the closed
  symbols and component 2 is represented by the open symbols.  The
  different kinds of collisions are: \FilledSquare \ intra-component,
  particle-particle ($\langle \Delta E_1 \rangle_1$ and $\langle
  \Delta E_2 \rangle_2$), \FilledCircle \ inter-component,
  particle-particle ($\langle \Delta E_1 \rangle_2$ and $\langle
  \Delta E_2 \rangle_1$), \FilledDiamondshape \ particle-wall, and
  \FilledTriangleUp \ particle-base. The solid (dashed) lines are
  the energy loss for component 1 (component 2) as calculated by
  Equation \ref{eqn:enloss1} for intra-component (same shade as the
  squares) and inter-component (same shade as the circles),
  particle-particle collisions. \label{fig:sr:energy}}
\end{figure}

Figures \ref{fig:sr:energy}a and \ref{fig:sr:energy}b show the average
energy changes that particles of components 1 and 2 experience as a
result of collision.  Collisions with the wall cause a loss of energy
for both components while collision with base increase the energy of
the particles for all size ratios (Figure \ref{fig:sr:energy}a).  The
collisions between particles, however show different trends than seen
in the mass ratio study.  As seen in Figure \ref{fig:sr:energy}b, the
inter-component collisions do not result in an
energy loss for component 1 for all size ratios.  The smaller the size
ratio, the greater the amount of energy injected per collision into
component 1 from collisions with component 2.  All particle-particle
collisions decrease the energy of component 2. The energy loss per
collision for for both inter-component and intra-component,
particle-particle collisions increase as the size ratio decreases.
It is interesting to note that the energy loss due to both
inter-component and intra-component collisions are the same for
component 2 at the smallest size ratio.  The energy changes for all
but the smallest size ratio agree well with those predicted by
Equation \ref{eqn:enloss1}.  The smallest size ratio shows a large
deviation between the energy loss predicted by the equations and that
determined from the simulation.  One reason for the discrepancy is the
assumption of Maxwell-Boltzmann velocities used in determining the
equations.  The differences in the distribution of energy within each
component and between the two components that we observe here will
affect the packing fraction and temperature for these systems.
 
\subsubsection{Packing Fraction \label{sec:sr:pf}} 
 
\begin{figure} 
\includegraphics[width=3.375in, keepaspectratio=true]{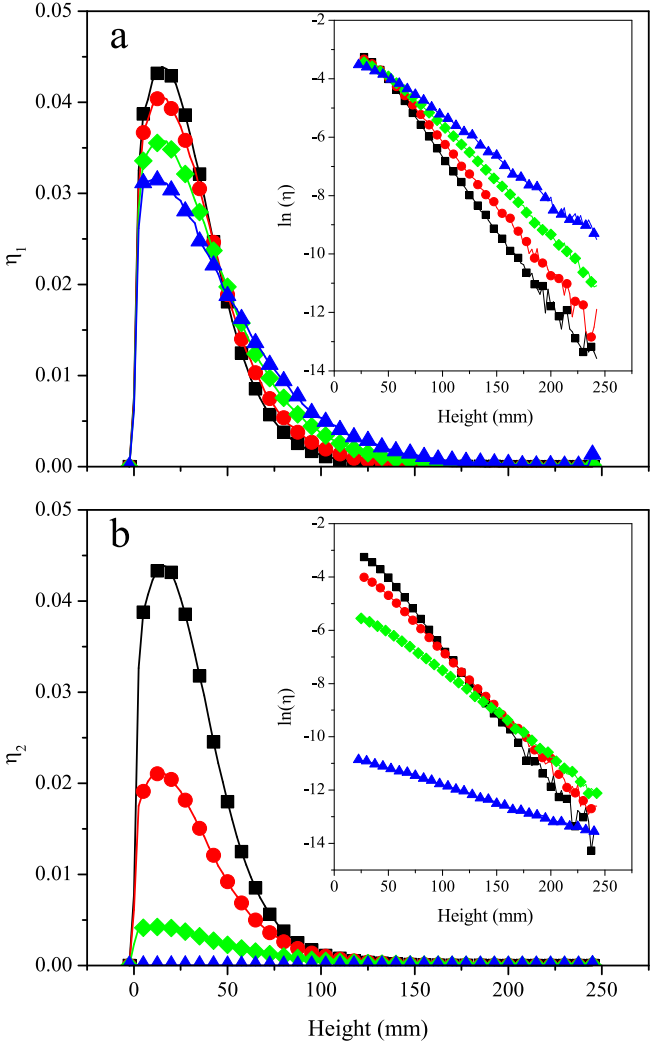}
\caption{The packing fraction of (a) component 1 and (b) component
  2 for size ratios of \FilledSquare~$d_2/d_1 = 1.0$,
\FilledCircle~$d_2/d_1 = 0.8$, \FilledDiamondshape~$d_2/d_1 = 0.5$,
and \FilledTriangleUp~$d_2/d_1 = 0.1$. All other simulation parameters
  are as given in the text.  The inset shows the natural log
of the trailing edge of the packing fraction for each system.
\label{fig:sr:pf}} 
\end{figure}

Figure \ref{fig:sr:pf} shows the effect of the size ratio on the 
packing fraction of the system.  We can see that component 1 reaches
higher altitudes as the diameter of component 2 decreases without any
noticeable shift in the position of the maximum in packing fraction.
This implies that the particles of component 1 expand through the
system as the size ratio decreases.  There are two possible causes of
the increase in the tail of the packing fraction at large heights.
First, the particles of component 1 are able to retain more energy
because of a decrease in the collision rate between particles as the
size ratio decreases.  The decrease in the collision rates is the
result of a decrease in the total excluded volume of the system as the
size of component 2 is reduced (deduced from Figures \ref{fig:sr:pf}a
and \ref{fig:sr:pf}b). The change in the total excluded volume of the
system reduces the amount of energy lost to particle-particle
collisions.  The other cause of the increased altitude is the energy
gain that comes from collisions with component 2.  Since energy can be
gained at positions above the base, the particles will be able
to travel to higher altitudes in the system.

Figure \ref{fig:sr:pf}b shows the effect of the size ratio on 
component 2.  We can see an overall reduction in the packing 
fraction of component 2 as the particle size is reduced.  This occurs
because the number of  particles of component 2 is held fixed as the
size is decreased.  We  also see that the particles of component 2 are
able to reach  greater altitudes in the system as the size ratio
decreases (see the  insert in Figure \ref{fig:sr:pf}b).  The increase
in the altitude is  the result of the decrease in the number of
collisions discussed above. 
 
\subsubsection{Temperature \label{sec:sr:t}} 
 
\begin{figure} 
\includegraphics[width=3.375in, keepaspectratio=true]{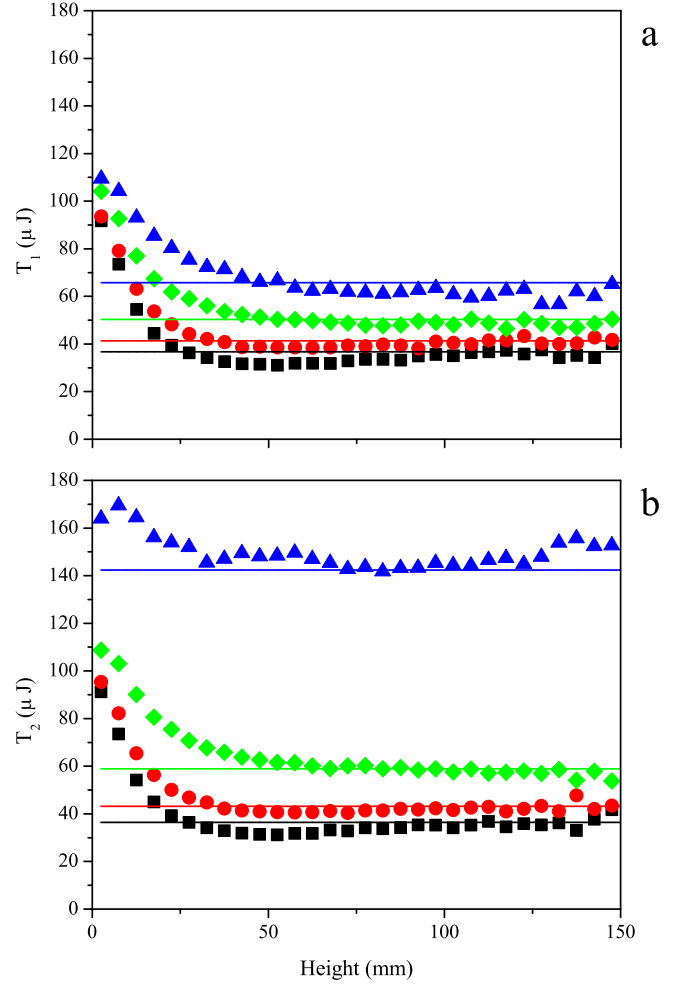}
\caption{The temperature in the z-direction of (a) component 1 and
(b) component 2 for size ratios of \FilledSquare~$d_2/d_1 = 1.0$,
  \FilledCircle~$d_2/d_1 = 0.8$, \FilledDiamondshape~$d_2/d_1 = 0.5$,
  and \FilledTriangleUp~$d_2/d_1 = 0.1$.  The temperature calculated
  from Equation \ref{eqn:slope} is shown as a line corresponding to
  the data points of the same shade.  \label{fig:sr:t}}
\end{figure}

Figure \ref{fig:sr:t}a shows the z-component of the granular
temperature as a function of height for component 1 for the four size
ratios.  Figure \ref{fig:sr:t}b shows the same for component 2 of the
mixture.  A minimum is again observed in the temperatures of each
system for each component.  The other general features of the
temperature profiles in the x- and y-directions, while not shown here,
are the same as seen and discussed in connection with Figure
\ref{fig:wp:t2}.  It can be easily seen in Figure \ref{fig:sr:t} that
the temperature of both the large and small particles increase as the
size ratio decreases.  This increase in temperature results from a
decrease in the energy lost due to particle-particle collisions.  The
reduction in the collision rate as the size ratio decreases can be
observed in Table \ref{tab:sr:col}.  It is interesting that the energy
loss due to intra-component collisions increases for component 2 as
the collision rate for intra-component collisions decreases (Figure
\ref{fig:sr:energy}b).  In addition, there is a discrepancy between
the theory and the simulation results observed in Figure
\ref{fig:sr:energy} at small size ratios that should be noted.  The
deviations are the result of the systems being dominated by collisions
with the wall and not particle-particle collisions (see Table
\ref{tab:sr:col}), as assumed in the theory (see the Appendix).
 
%\begin{turnpage}
\begin{table*} 
\caption{Collision rates ($s^{-1}$) for the different kinds of collisions for 
components 1 and 2. \label{tab:sr:col}} \begin{ruledtabular} 
\begin{tabular}{|c|cccc|cccc|} 
&\multicolumn{4}{c|}{Component 1}&\multicolumn{4}{c|}{Component 2}\\ 
\hline Size 
Ratio&intra-component&inter-component&Wall&Base&intra-component& 
inter-component&Wall&Base\\ \hline 
1.0&272.8&546.2&181.2&78.8&272.1&546.2&195.1&78.8\\ 
0.8&252.4&412.0&278.5&74.7&164.1&412.0&381.2&77.1\\ 
0.5&230.0&248.1&503.0&70.3&55.2&248.1&2587.4&71.9\\ 
0.1&208.9&72.5&649.4&68.1&7.75&72.5&136270&54.4\\ 
\end{tabular} 
\end{ruledtabular} 
\end{table*} 
%\end{turnpage}
 
We used Equation \ref{eqn:slope} and the high altitude tails of the
packing fractions shown in the inserts in Figures \ref{fig:sr:pf}a and
\ref{fig:sr:pf}b to calculate a temperature.  The temperatures
calculated in this manner are also plotted in Figure
\ref{fig:sr:t}. The temperature calculated from Equation
\ref{eqn:slope} coincides with the temperature in the z-direction at
large heights, just as in the case of the mass ratio. The agreement
between the two temperatures is very good for both components for each
size ratio, except the smallest. This may indicate that the packing
fraction is not a very sensitive measure of the temperature when the
difference in size between the components is large.  The difference in
the two temperatures may also be indicative of the changes in the
dominant processes in the energy distribution. For example, the
smallest size ratio is not dominated by particle-particle collisions
like the other systems.  The edge effects associated with the walls of
the cylinder become more important, but they are not considered in the
theory used to obtain Equation \ref{eqn:slope}.
 
\section{Conclusions \label{sec:con}} 

We have shown that, like the single component system \cite{TV2002},
the simulation results reproduce the phenomena observed in
experimental studies.  Specifically, the experimentally observed
\cite{WP2002} flow pattern, radial dependence of the packing fraction
and temperature of the two components are qualitatively reproduced by
the simulation. 

There are a number of possible reasons for the quantitative
differences between the simulation and experiments. For example, for
each component the simulated particles are identical spheres, while in
the experimental system there is a small distribution of shape, size,
and mass.  The simulation also assumes that the particles are
frictionless with constant restitution coefficients, which is not the
case for ballotini glass spheres used in the experiments.  The
saw-tooth wave form of the vibrating base is an idealization of the
sinusoidal vibration of the experiment.  It is thought, however, that
this does not have a large influence on the system behavior
\cite{MB1997}.  In any case, we feel confident that our model captures
the key physical aspects of the experimental system and can therefore
be used to study the influence of various system parameters.

We also studied the effects of mass and size ratio in this paper.
Generally, we observed that changes in either ratio do not result in
any segregation of the particles.  The lighter particles attain
greater heights than the heavier particles in the mass ratio studies.
For both components the rate of decay of the packing fraction at large
heights is a good indication of the granular temperature in that
region \cite{BR-MM2001,WHJ1995}.

As the mass ratio decreases, the overall temperature of the system
decreases.  This is consistent with the lower amount of energy gained
by the lighter component from the vibrating base and an overall
lowering of the efficiency of energy transfer as the mass ratio
decreases.  The packing fraction and temperature of the individual
components, however, have a non-trivial dependence on the mass ratio.
For the heavier component, both these quantities exhibit a minimum as
the mass ratio decreases.  We also observed that the energy changes
due to inter-component collisions exhibit extrema for each component
as the mass ratio decreases.  Specifically, the heavier component
shows a minimum, whereas the lighter component shows a maximum in the
energy change.  The lighter component actually gains energy from
collisions with the other component around a mass ratio of
$m_2/m_1=0.5$.  The energy changes due to intra-component collisions
for both components, however, are negative for all mass ratios.  The
energy loss due to intra-particle collisions increases for the heavy
component as the mass ratio decreases, while it decreases for the
lighter component.  For comparison, we developed an approximate theory
to calculate the energy changes for particle-particle collisions by
assuming that the particle velocities follow a Maxwell-Boltzmann
distribution with a component specific temperature.  While this
assumption is not strictly correct given the inhomogeneities in
density and temperature, the theory is qualitatively accurate for the
inter-component collisions and in near quantitative agreement with the
simulation for the intra-component collisions.

As the size ratio decreases at constant mass ratio, the overall
temperature of the system increases and the total packing fraction
decreases. The particle-particle and particle-boundary collision rates
decrease and increase, respectively, as the size ratio decreased.  At
the same time, the larger particles begin to gain energy from
collisions with the smaller component. The approximate theory of the
energy changes is again able to reproduce qualitatively the observed
trends.  However, the agreement between the theory and simulation
results worsens as the size ratio decreases.

Wildman and Parker \cite{WP2002} observed a decrease in the
temperature as they decreased the ratio of the number of large to
small ballotini spheres. Our results show that this effect is
dominated by the difference in mass of the two components, and not the
difference in size.

\begin{acknowledgments}
We thank Ricky Wildman for useful discussions and the National
Science Foundation (CHE-9814236) for financial support.
\end{acknowledgments}
 
\appendix* 
\section{Energy dissipation equations \label{app}} 
The energy change of 
particle $i$ of component $\alpha$ resulting from collision with
particle $j$ of component $\beta$ is
\begin{eqnarray}
\label{aeqn:enlossi} 
2 m_\alpha \Delta E_{\alpha,i} = \mu^2 \left( 1+c \right)^2 \left(
\mathbf{v_{ij}} \cdot \mathbf{\hat{n}} \right)^2  
 - m_\alpha \mu \left( 1+c \right) \nonumber \\ 
\times \left[ \left( \mathbf{v_{\alpha,i}} \cdot 
  \mathbf{\hat{n}} \right)^2 - \left( \mathbf{v_{\beta,j}} \cdot 
  \mathbf{\hat{n}} \right)^2 + \left( \mathbf{v_{ij}} \cdot 
\mathbf{\hat{n}} \right)^2 \right] 
\end{eqnarray} 
where $\mathbf{v_{ij}} = \mathbf{v_{\alpha,i}} -
\mathbf{v_{\beta,j}}$ and $\mu = m_\alpha\, m_\beta/(m_\alpha+m_\beta)$.  
The total energy loss due to the collision is simply $ \Delta 
E_{total} = \Delta E_{\alpha,i} + \Delta E_{\beta,j}$, or 
\begin{equation} 
\label{aeqn:enlosst} 
\Delta E_{total}=- \frac{1}{2} \mu \left(1-c^2\right) \left( 
\mathbf{v_{ij}} \cdot \mathbf{\hat{n}} \right)^2 
\end{equation} 
These equations are exact for each collision that occurs in the
system.  In order to determine the average values, however, some
assumptions must be made.  Specifically, we assume that the velocities
of each component are described by a Maxwell-Boltzmann distribution
that is homogeneous, isotropic and characterized by a
component-specific temperature
$T_\alpha$:
\begin{equation}
f \left(\mathbf{v_{\alpha,i}} \right)d\mathbf{v_{\alpha,i}}=\left(\frac{2\pi
m_\alpha}{k_BT_\alpha}\right)^{3/2}\exp\left(-
\frac{m_\alpha v_{\alpha,i}^2}{2 k_BT_\alpha}\right)d\mathbf{v_{\alpha,i}}
\end{equation}
In order to average over the velocities
$\mathbf{v_{\alpha,i}}$, $\mathbf{v_{\beta,j}}$, and $\mathbf{v_{ij}}$
appearing in Equations \ref{aeqn:enlossi} and \ref{aeqn:enlosst}, we introduce
center-of-mass, $\mathbf{v_c}$, and relative $\mathbf{v_r}$
velocities: 
\begin{eqnarray} 
\mathbf{v_{\alpha,i}} =
\frac{m_\beta/T_\beta}{m_\alpha/T_\alpha+m_\beta/T_\beta} \mathbf{v_r}
+ \mathbf{v_c}\\  
\mathbf{v_{\beta,j}} =
-\frac{m_\alpha/T_\alpha}{m_\alpha/T_\alpha+m_\beta/T_\beta}
\mathbf{v_r} + \mathbf{v_c}\\ 
\mathbf{v_{ij}} = \mathbf{v_{\alpha,i}}-\mathbf{v_{\beta,j}} = \mathbf{v_r} 
\end{eqnarray} 
In this coordinate system the energy change for particle $i$ becomes 
\begin{eqnarray} 
\label{aeqn:enlossi2} 
2 m_\alpha \Delta E_{\alpha,i} = && \mu^2 \left( 1+c \right)^2 \left(
\mathbf{v_r} \cdot \mathbf{\hat{n}} \right)^2 \nonumber \\ 
&& - 2 T_\alpha \mu' \mu \left( 1+c 
  \right) \left( \mathbf{v_r} \cdot 
\mathbf{\hat{n}} \right)^2 \\  
&& - 2 m_\alpha \mu \left( 1+c \right)  \left( 
\mathbf{v_r} \cdot \mathbf{\hat{n}} \right) \left( \mathbf{v_c} \cdot 
\mathbf{\hat{n}} \right) \nonumber 
\end{eqnarray} 
where  
\begin{equation} 
\mu'=\frac{(m_\alpha/T_\alpha)(m_\beta/T_\beta)}{m_\alpha/T_\alpha +
  m_\beta/T_\beta}
\end{equation} 
The total energy loss (shown in Equation \ref{aeqn:enlosst}) becomes
\begin{equation}
\label{aeqn:enlosst2}
\Delta E_{total} = - \frac{1}{2} \mu \left( 1-c^2 \right) \left(
\mathbf{v_r} \cdot \mathbf{\hat{n}} \right)^2
\end{equation}
in the new coordinate system.

It is now necessary to average Equations \ref{aeqn:enlossi2} and
\ref{aeqn:enlosst2} over the fraction of collisions between a particle
of component $\alpha$ with a particle of component $\beta$ with a relative
velocity between $v_r$ and $v_r+dv_r$. Straightforward modification of
the standard kinetic theory result \cite{McQuarrie} gives:
\begin{equation}
p \left(v_r \right)dv_r=\frac{1}{2}\left(\frac{\mu'}{k_B}\right)^2 v_r^3 
\exp\left(-\frac{\mu'v_r^2}{2k_B}\right)dv_r
\end{equation}
We then compute
\begin{equation}
\langle \left( \mathbf{v_r} \cdot \mathbf{\hat{n}} \right)^2\rangle=
\int_0^{\infty}dv_rp(v_r)\int_0^{\sigma}db\,
h(b)\left( \mathbf{v_r} \cdot \mathbf{\hat{n}} \right)^2
\end{equation}
where $h(b)db=(2b/\sigma^2)db$ is the probability that the impact
parameter lies between $b$ and $b+db$. Substituting 
$\left( \mathbf{v_r} \cdot \mathbf{\hat{n}}
\right)^2=v_r^2(1-(b/r)^2)$ and evaluating the
integrals leads to 
\begin{equation}
\langle \left( \mathbf{v_r} \cdot \mathbf{\hat{n}}
\right)^2 \rangle=\frac{2k_B}{\mu'}
\end{equation}
Using this result and the fact that
$\langle \left( 
\mathbf{v_r} \cdot \mathbf{\hat{n}} \right) \left( \mathbf{v_c} \cdot 
\mathbf{\hat{n}} \right) \rangle =0$ gives Equations \ref{eqn:enloss1} and 
\ref{eqn:enloss2}.

\bibliography{granmat} 

\begin{thebibliography}{32}
\expandafter\ifx\csname natexlab\endcsname\relax\def\natexlab#1{#1}\fi
\expandafter\ifx\csname bibnamefont\endcsname\relax
  \def\bibnamefont#1{#1}\fi
\expandafter\ifx\csname bibfnamefont\endcsname\relax
  \def\bibfnamefont#1{#1}\fi
\expandafter\ifx\csname citenamefont\endcsname\relax
  \def\citenamefont#1{#1}\fi
\expandafter\ifx\csname url\endcsname\relax
  \def\url#1{\texttt{#1}}\fi
\expandafter\ifx\csname urlprefix\endcsname\relax\def\urlprefix{URL }\fi
\providecommand{\bibinfo}[2]{#2}
\providecommand{\eprint}[2][]{\url{#2}}

\bibitem[{\citenamefont{Wildman and Parker}(2002)}]{WP2002}
\bibinfo{author}{\bibfnamefont{R.~D.} \bibnamefont{Wildman}} \bibnamefont{and}
  \bibinfo{author}{\bibfnamefont{D.~J.} \bibnamefont{Parker}},
  \bibinfo{journal}{Phys. Rev. Lett.} \textbf{\bibinfo{volume}{88}},
  \bibinfo{pages}{064301} (\bibinfo{year}{2002}).

\bibitem[{\citenamefont{Douday et~al.}(1989)\citenamefont{Douday, Fauve, and
  Laroche}}]{DFL1989}
\bibinfo{author}{\bibfnamefont{S.}~\bibnamefont{Douday}},
  \bibinfo{author}{\bibfnamefont{S.}~\bibnamefont{Fauve}}, \bibnamefont{and}
  \bibinfo{author}{\bibfnamefont{C.}~\bibnamefont{Laroche}},
  \bibinfo{journal}{Europhys. Lett.} \textbf{\bibinfo{volume}{8}},
  \bibinfo{pages}{621} (\bibinfo{year}{1989}).

\bibitem[{\citenamefont{Pak and Behringer}(1993)}]{PB1993}
\bibinfo{author}{\bibfnamefont{H.~K.} \bibnamefont{Pak}} \bibnamefont{and}
  \bibinfo{author}{\bibfnamefont{R.~P.} \bibnamefont{Behringer}},
  \bibinfo{journal}{Phys. Rev. Lett.} \textbf{\bibinfo{volume}{71}},
  \bibinfo{pages}{1832} (\bibinfo{year}{1993}).

\bibitem[{\citenamefont{Melo et~al.}(1995)\citenamefont{Melo, Umbanhowar, and
  Swinney}}]{MUS1995}
\bibinfo{author}{\bibfnamefont{F.}~\bibnamefont{Melo}},
  \bibinfo{author}{\bibfnamefont{P.~B.} \bibnamefont{Umbanhowar}},
  \bibnamefont{and} \bibinfo{author}{\bibfnamefont{H.~L.}
  \bibnamefont{Swinney}}, \bibinfo{journal}{Phys. Rev. Lett.}
  \textbf{\bibinfo{volume}{75}}, \bibinfo{pages}{3838} (\bibinfo{year}{1995}).

\bibitem[{\citenamefont{Umbanhowar et~al.}(1996)\citenamefont{Umbanhowar, Melo,
  and Swinney}}]{UMS1996}
\bibinfo{author}{\bibfnamefont{P.~B.} \bibnamefont{Umbanhowar}},
  \bibinfo{author}{\bibfnamefont{F.}~\bibnamefont{Melo}}, \bibnamefont{and}
  \bibinfo{author}{\bibfnamefont{H.~L.} \bibnamefont{Swinney}},
  \bibinfo{journal}{Nature (London)} \textbf{\bibinfo{volume}{382}},
  \bibinfo{pages}{793} (\bibinfo{year}{1996}).

\bibitem[{\citenamefont{Evesque and Rajchenbach}(1989)}]{ER1989}
\bibinfo{author}{\bibfnamefont{P.}~\bibnamefont{Evesque}} \bibnamefont{and}
  \bibinfo{author}{\bibfnamefont{J.}~\bibnamefont{Rajchenbach}},
  \bibinfo{journal}{Phys. Rev. Lett.} \textbf{\bibinfo{volume}{62}},
  \bibinfo{pages}{44} (\bibinfo{year}{1989}).

\bibitem[{\citenamefont{Ehrichs et~al.}(1995)\citenamefont{Ehrichs, Jaeger,
  Karczmar, Knight, Kuperman, and Nagel}}]{EJKKKN1995}
\bibinfo{author}{\bibfnamefont{E.~E.} \bibnamefont{Ehrichs}},
  \bibinfo{author}{\bibfnamefont{H.~M.} \bibnamefont{Jaeger}},
  \bibinfo{author}{\bibfnamefont{G.~S.} \bibnamefont{Karczmar}},
  \bibinfo{author}{\bibfnamefont{J.~B.} \bibnamefont{Knight}},
  \bibinfo{author}{\bibfnamefont{V.~Y.} \bibnamefont{Kuperman}},
  \bibnamefont{and} \bibinfo{author}{\bibfnamefont{S.~R.} \bibnamefont{Nagel}},
  \bibinfo{journal}{Science} \textbf{\bibinfo{volume}{267}},
  \bibinfo{pages}{1632} (\bibinfo{year}{1995}).

\bibitem[{\citenamefont{Rosato et~al.}(1987)\citenamefont{Rosato, Shandburg,
  Prinz, and Swendsen}}]{RSPS1987}
\bibinfo{author}{\bibfnamefont{A.}~\bibnamefont{Rosato}},
  \bibinfo{author}{\bibfnamefont{K.~J.} \bibnamefont{Shandburg}},
  \bibinfo{author}{\bibfnamefont{F.}~\bibnamefont{Prinz}}, \bibnamefont{and}
  \bibinfo{author}{\bibfnamefont{R.~H.} \bibnamefont{Swendsen}},
  \bibinfo{journal}{Phys. Rev. Lett.} \textbf{\bibinfo{volume}{58}},
  \bibinfo{pages}{1038} (\bibinfo{year}{1987}).

\bibitem[{\citenamefont{Losert et~al.}(1999)\citenamefont{Losert, Cooper,
  Delour, Kudrolli, and Gollub}}]{L1999}
\bibinfo{author}{\bibfnamefont{W.}~\bibnamefont{Losert}},
  \bibinfo{author}{\bibfnamefont{D.~G.~W.} \bibnamefont{Cooper}},
  \bibinfo{author}{\bibfnamefont{J.}~\bibnamefont{Delour}},
  \bibinfo{author}{\bibfnamefont{A.}~\bibnamefont{Kudrolli}}, \bibnamefont{and}
  \bibinfo{author}{\bibfnamefont{J.~P.} \bibnamefont{Gollub}},
  \bibinfo{journal}{Chaos} \textbf{\bibinfo{volume}{9}}, \bibinfo{pages}{682}
  (\bibinfo{year}{1999}).

\bibitem[{\citenamefont{McNamara and Luding}(1998)}]{ML1998}
\bibinfo{author}{\bibfnamefont{S.}~\bibnamefont{McNamara}} \bibnamefont{and}
  \bibinfo{author}{\bibfnamefont{S.}~\bibnamefont{Luding}},
  \bibinfo{journal}{Phys. Rev. E} \textbf{\bibinfo{volume}{58}},
  \bibinfo{pages}{2247} (\bibinfo{year}{1998}).

\bibitem[{\citenamefont{Garzo and Dufty}(1999)}]{GD1999}
\bibinfo{author}{\bibfnamefont{V.}~\bibnamefont{Garzo}} \bibnamefont{and}
  \bibinfo{author}{\bibfnamefont{J.}~\bibnamefont{Dufty}},
  \bibinfo{journal}{Phys. Rev. E} \textbf{\bibinfo{volume}{60}},
  \bibinfo{pages}{5706} (\bibinfo{year}{1999}).

\bibitem[{\citenamefont{Dahl et~al.}(2002)\citenamefont{Dahl, Hrenya, Garzo,
  and Dufty}}]{DHGD}
\bibinfo{author}{\bibfnamefont{S.~R.} \bibnamefont{Dahl}},
  \bibinfo{author}{\bibfnamefont{C.~M.} \bibnamefont{Hrenya}},
  \bibinfo{author}{\bibfnamefont{V.}~\bibnamefont{Garzo}}, \bibnamefont{and}
  \bibinfo{author}{\bibfnamefont{J.~W.} \bibnamefont{Dufty}},
  \emph{\bibinfo{title}{Kinetic temperatures for a granular mixture}}
  (\bibinfo{year}{2002}), \eprint{cond-mat/0205413}.

\bibitem[{\citenamefont{Marconi and Puglisi}(2002{\natexlab{a}})}]{MP2002}
\bibinfo{author}{\bibfnamefont{U.~M.~B.} \bibnamefont{Marconi}}
  \bibnamefont{and} \bibinfo{author}{\bibfnamefont{A.}~\bibnamefont{Puglisi}},
  \bibinfo{journal}{Phys. Rev. E} \textbf{\bibinfo{volume}{65}},
  \bibinfo{pages}{051305} (\bibinfo{year}{2002}{\natexlab{a}}).

\bibitem[{\citenamefont{Marconi and Puglisi}(2002{\natexlab{b}})}]{MP2002a}
\bibinfo{author}{\bibfnamefont{U.~M.~B.} \bibnamefont{Marconi}}
  \bibnamefont{and} \bibinfo{author}{\bibfnamefont{A.}~\bibnamefont{Puglisi}},
  \bibinfo{journal}{Phys. Rev. E} \textbf{\bibinfo{volume}{66}},
  \bibinfo{pages}{011301} (\bibinfo{year}{2002}{\natexlab{b}}).

\bibitem[{\citenamefont{Barrat and Trizac}(2002{\natexlab{a}})}]{BT2002}
\bibinfo{author}{\bibfnamefont{A.}~\bibnamefont{Barrat}} \bibnamefont{and}
  \bibinfo{author}{\bibfnamefont{E.}~\bibnamefont{Trizac}},
  \bibinfo{journal}{Granul. Matter} \textbf{\bibinfo{volume}{4}},
  \bibinfo{pages}{57} (\bibinfo{year}{2002}{\natexlab{a}}).

\bibitem[{\citenamefont{Barrat and Trizac}(2002{\natexlab{b}})}]{BTa}
\bibinfo{author}{\bibfnamefont{A.}~\bibnamefont{Barrat}} \bibnamefont{and}
  \bibinfo{author}{\bibfnamefont{E.}~\bibnamefont{Trizac}},
  \emph{\bibinfo{title}{Molecular dynamics simulation of vibrated granular
  gases}} (\bibinfo{year}{2002}{\natexlab{b}}), \eprint{cond-mat/0207267}.

\bibitem[{\citenamefont{Feitosa and Menon}(2001)}]{FM}
\bibinfo{author}{\bibfnamefont{K.}~\bibnamefont{Feitosa}} \bibnamefont{and}
  \bibinfo{author}{\bibfnamefont{N.}~\bibnamefont{Menon}},
  \emph{\bibinfo{title}{Breakdown of energy equipartition in a 2d binary
  vibrated granular gas}} (\bibinfo{year}{2001}), \eprint{cond-mat/0111391}.

\bibitem[{\citenamefont{Talbot and Viot}(2002)}]{TV2002}
\bibinfo{author}{\bibfnamefont{J.}~\bibnamefont{Talbot}} \bibnamefont{and}
  \bibinfo{author}{\bibfnamefont{P.}~\bibnamefont{Viot}},
  \bibinfo{journal}{Phys. Rev. Lett.} \textbf{\bibinfo{volume}{89}},
  \bibinfo{pages}{064301} (\bibinfo{year}{2002}).

\bibitem[{\citenamefont{Paolotti et~al.}()\citenamefont{Paolotti, Cattuto,
  Marconi, and Puglisi}}]{PCMP}
\bibinfo{author}{\bibfnamefont{D.}~\bibnamefont{Paolotti}},
  \bibinfo{author}{\bibfnamefont{C.}~\bibnamefont{Cattuto}},
  \bibinfo{author}{\bibfnamefont{U.~M.~B.} \bibnamefont{Marconi}},
  \bibnamefont{and} \bibinfo{author}{\bibfnamefont{A.}~\bibnamefont{Puglisi}},
  \eprint{cond-mat/0207601}.

\bibitem[{\citenamefont{Barrat and Trizac}(2001)}]{BTb}
\bibinfo{author}{\bibfnamefont{A.}~\bibnamefont{Barrat}} \bibnamefont{and}
  \bibinfo{author}{\bibfnamefont{E.}~\bibnamefont{Trizac}},
  \emph{\bibinfo{title}{Heated granular fluids: the random restitution
  coefficient approach}} (\bibinfo{year}{2001}), \eprint{cond-mat/0207267}.

\bibitem[{\citenamefont{Clelland and Hrenya}(2002)}]{CH2002}
\bibinfo{author}{\bibfnamefont{R.}~\bibnamefont{Clelland}} \bibnamefont{and}
  \bibinfo{author}{\bibfnamefont{C.~M.} \bibnamefont{Hrenya}},
  \bibinfo{journal}{Phys. Rev. E} \textbf{\bibinfo{volume}{65}},
  \bibinfo{pages}{031301} (\bibinfo{year}{2002}).

\bibitem[{\citenamefont{Barrat et~al.}(2001)\citenamefont{Barrat, Trizac, and
  Fuchs}}]{BTF2001}
\bibinfo{author}{\bibfnamefont{A.}~\bibnamefont{Barrat}},
  \bibinfo{author}{\bibfnamefont{E.}~\bibnamefont{Trizac}}, \bibnamefont{and}
  \bibinfo{author}{\bibfnamefont{J.~N.} \bibnamefont{Fuchs}},
  \bibinfo{journal}{Eur. Phys. J. E} \textbf{\bibinfo{volume}{5}},
  \bibinfo{pages}{161} (\bibinfo{year}{2001}).

\bibitem[{\citenamefont{Brilliantov et~al.}(1996)\citenamefont{Brilliantov,
  Spahn, Hertzsch, and P{\"o}schel}}]{BSHP1996}
\bibinfo{author}{\bibfnamefont{N.~V.} \bibnamefont{Brilliantov}},
  \bibinfo{author}{\bibfnamefont{F.}~\bibnamefont{Spahn}},
  \bibinfo{author}{\bibfnamefont{J.~M.} \bibnamefont{Hertzsch}},
  \bibnamefont{and}
  \bibinfo{author}{\bibfnamefont{T.}~\bibnamefont{P{\"o}schel}},
  \bibinfo{journal}{Phys. Rev. E} \textbf{\bibinfo{volume}{53}},
  \bibinfo{pages}{5382} (\bibinfo{year}{1996}).

\bibitem[{\citenamefont{Goldman et~al.}(1998)\citenamefont{Goldman, Shattuck,
  Bizon, McCormick, Swift, and Swinney}}]{GSBMSS1998}
\bibinfo{author}{\bibfnamefont{D.}~\bibnamefont{Goldman}},
  \bibinfo{author}{\bibfnamefont{M.~D.} \bibnamefont{Shattuck}},
  \bibinfo{author}{\bibfnamefont{C.}~\bibnamefont{Bizon}},
  \bibinfo{author}{\bibfnamefont{W.~D.} \bibnamefont{McCormick}},
  \bibinfo{author}{\bibfnamefont{J.~B.} \bibnamefont{Swift}}, \bibnamefont{and}
  \bibinfo{author}{\bibfnamefont{H.~L.} \bibnamefont{Swinney}},
  \bibinfo{journal}{Phys. Rev. E} p. \bibinfo{pages}{4831}
  (\bibinfo{year}{1998}).

\bibitem[{\citenamefont{Luding and McNamara}(1998)}]{LM}
\bibinfo{author}{\bibfnamefont{S.}~\bibnamefont{Luding}} \bibnamefont{and}
  \bibinfo{author}{\bibfnamefont{S.}~\bibnamefont{McNamara}},
  \emph{\bibinfo{title}{How to handle the inelastic collapse of a dissipative
  hard-sphere gas with the {TC} model}} (\bibinfo{year}{1998}),
  \eprint{cond-mat/9810009}.

\bibitem[{\citenamefont{Wildman
  et~al.}(2001{\natexlab{a}})\citenamefont{Wildman, Huntley, and
  Parker}}]{WHP2001}
\bibinfo{author}{\bibfnamefont{R.~D.} \bibnamefont{Wildman}},
  \bibinfo{author}{\bibfnamefont{J.~M.} \bibnamefont{Huntley}},
  \bibnamefont{and} \bibinfo{author}{\bibfnamefont{D.~J.}
  \bibnamefont{Parker}}, \bibinfo{journal}{Phys. Rev. Lett.}
  \textbf{\bibinfo{volume}{86}}, \bibinfo{pages}{3304}
  (\bibinfo{year}{2001}{\natexlab{a}}).

\bibitem[{\citenamefont{Brey et~al.}(2001)\citenamefont{Brey, Ruiz-Montero, and
  Moreno}}]{BR-MM2001}
\bibinfo{author}{\bibfnamefont{J.~J.} \bibnamefont{Brey}},
  \bibinfo{author}{\bibfnamefont{M.~J.} \bibnamefont{Ruiz-Montero}},
  \bibnamefont{and} \bibinfo{author}{\bibfnamefont{F.}~\bibnamefont{Moreno}},
  \bibinfo{journal}{Phys. Rev. E} \textbf{\bibinfo{volume}{63}},
  \bibinfo{pages}{061305} (\bibinfo{year}{2001}).

\bibitem[{\citenamefont{Wildman
  et~al.}(2001{\natexlab{b}})\citenamefont{Wildman, Huntley, and
  Parker}}]{WHP2001a}
\bibinfo{author}{\bibfnamefont{R.~D.} \bibnamefont{Wildman}},
  \bibinfo{author}{\bibfnamefont{J.~M.} \bibnamefont{Huntley}},
  \bibnamefont{and} \bibinfo{author}{\bibfnamefont{D.~J.}
  \bibnamefont{Parker}}, \bibinfo{journal}{Phys. Rev. E}
  \textbf{\bibinfo{volume}{63}}, \bibinfo{pages}{061311}
  (\bibinfo{year}{2001}{\natexlab{b}}).

\bibitem[{\citenamefont{Ramirez and Soto}(2002)}]{RS}
\bibinfo{author}{\bibfnamefont{R.}~\bibnamefont{Ramirez}} \bibnamefont{and}
  \bibinfo{author}{\bibfnamefont{R.}~\bibnamefont{Soto}},
  \emph{\bibinfo{title}{Temperature inversion in granular fluids under
  gravity}} (\bibinfo{year}{2002}), \eprint{cond-mat/0210471}.

\bibitem[{\citenamefont{Warr et~al.}(1995)\citenamefont{Warr, Huntley, and
  Jacques}}]{WHJ1995}
\bibinfo{author}{\bibfnamefont{S.}~\bibnamefont{Warr}},
  \bibinfo{author}{\bibfnamefont{J.~M.} \bibnamefont{Huntley}},
  \bibnamefont{and} \bibinfo{author}{\bibfnamefont{G.}~\bibnamefont{Jacques}},
  \bibinfo{journal}{Phys. Rev. E} \textbf{\bibinfo{volume}{52}},
  \bibinfo{pages}{5583} (\bibinfo{year}{1995}).

\bibitem[{\citenamefont{McNamara and Barrat}(1997)}]{MB1997}
\bibinfo{author}{\bibfnamefont{S.}~\bibnamefont{McNamara}} \bibnamefont{and}
  \bibinfo{author}{\bibfnamefont{J.~L.} \bibnamefont{Barrat}},
  \bibinfo{journal}{Phys. Rev. E} \textbf{\bibinfo{volume}{55}},
  \bibinfo{pages}{7767} (\bibinfo{year}{1997}).

\bibitem[{\citenamefont{McQuarrie}(1976)}]{McQuarrie}
\bibinfo{author}{\bibfnamefont{D.~A.} \bibnamefont{McQuarrie}},
  \emph{\bibinfo{title}{Statistical Mechanics}} (\bibinfo{publisher}{Harper and
  Row}, \bibinfo{year}{1976}), chap. \bibinfo{chapter}{7 and 16}.

\end{thebibliography}
 
\end{document}